\documentclass[twocolumn,hyperpdf,amsmath,amssymb,aps,prd,10pt,superscriptaddress,nofootinbib,noeprint,preprintnumbers,floatfix]{revtex4-1}

\usepackage{graphicx, color}

\usepackage[letterspace=-10]{microtype} 

\usepackage{bm, amsmath, amsfonts}
\usepackage{multirow, tabularx, dcolumn}

\usepackage[utf8]{inputenc} 
\usepackage{hyperref}

\definecolor{jlab_red}{RGB}{192,39,45}
\definecolor{jlab_orange}{RGB}{249,102,0}
\definecolor{jlab_blue}{RGB}{47,122,121}
\definecolor{jlab_green}{RGB}{65,125,10}

\hypersetup{%
pdftitle = {The quark mass dependence of elastic $\pi K$ scattering from QCD},
pdfsubject = {QCD},
pdfkeywords = {QCD, Hadron, Physics, Lattice, Meson, Scattering},
pdfauthor = {Hadron Spectrum Collaboration},
colorlinks = {true},
filecolor = {black},
linkcolor = {jlab_blue},
menucolor = {black},
citecolor = {jlab_green},
urlcolor = {jlab_green},
}{}

\begin{document}

\preprint{DAMTP-2019-13, JLAB-THY-19-2911}

\title{The quark-mass dependence of elastic $\pi K$ scattering from QCD}

\author{David~J.~Wilson}
\email{djwilson@maths.tcd.ie}
\affiliation{School of Mathematics, Trinity College, Dublin~2, Ireland}
\author{Ra\'ul~A.~Brice\~{n}o}
\email{rbriceno@jlab.org}
\affiliation{\lsstyle Thomas Jefferson National Accelerator Facility, 12000 Jefferson Avenue, Newport News, VA 23606, USA}
\affiliation{\lsstyle Department of Physics, Old Dominion University, Norfolk, VA 23529, USA}
\author{Jozef~J.~Dudek}
\email{dudek@jlab.org}
\affiliation{\lsstyle Thomas Jefferson National Accelerator Facility, 12000 Jefferson Avenue, Newport News, VA 23606, USA}
\affiliation{Department of Physics, College of William and Mary, Williamsburg, VA 23187, USA}
\author{Robert~G.~Edwards}
\email{edwards@jlab.org}
\affiliation{\lsstyle Thomas Jefferson National Accelerator Facility, 12000 Jefferson Avenue, Newport News, VA 23606, USA}
\author{Christopher~E.~Thomas}
\email{c.e.thomas@damtp.cam.ac.uk}
\affiliation{DAMTP, University of Cambridge, Centre for Mathematical Sciences, Wilberforce Road, Cambridge, CB3 0WA, UK}

\collaboration{for the Hadron Spectrum Collaboration}
\date{April 5, 2019}

\begin{abstract}
We present a determination of the isospin-$\frac{1}{2}$ elastic $\pi K$ scattering amplitudes in $S$ and $P$ partial waves using lattice Quantum Chromodynamics. The amplitudes, constrained for a large number of real-valued energy points, are obtained as a function of light-quark mass, corresponding to four pion masses between 200 and 400 MeV. 
Below the first inelastic threshold, the $P$-wave scattering amplitude is dominated by a single pole singularity
that evolves from being a stable bound-state at the highest quark mass into a narrow resonance that broadens as the pion and kaon masses are reduced.
As in experiment, the $S$-wave amplitude does not exhibit an obviously resonant behavior, but instead shows a slow rise from threshold, which is not inconsistent with the presence of a $\kappa$/$K_0^\star(700)$-like resonance at the considered quark masses. As has been found in analyses of experimental scattering data, simple analytic continuations into the complex energy plane of precisely-determined lattice QCD amplitudes on the real energy axis are not sufficient to model-independently determine the existence and properties of this state.
The spectra and amplitudes we present will serve as an input for increasingly elaborate amplitude analysis techniques that implement more of the analytic structure expected at complex energies.

\end{abstract}

\maketitle

\emph{Introduction} --- $\pi K$ scattering has a long history, which mirrors closely the $\pi\pi$ case, with the $P$-wave containing a clearly visible narrow resonance, the $K^\star(892)$ which partners the $\rho$, while the $S$-wave features only a slow rise with energy. Much of our experimental knowledge is derived from the classic kaon beam experiments~\cite{Estabrooks:1977xe, *Aston:1986jb, *Aston:1987ey, *Aston:1987ir} at SLAC where the dominance of pion exchange at small momentum transfers to proton targets was used to access an effective $\pi K$ initial state.

In a world where SU(3) flavor symmetry were exact, scattering amplitudes in isospin-$\frac{1}{2}$, isospin-1, and isospin-$0$ would all appear in an octet and have a common resonant content in each partial-wave. Empirically, these channels show strikingly different behavior in $S$-wave, indicating a strong breaking of the SU(3) flavor symmetry. How the experimental observations evolve towards the SU(3) symmetric theory with varying quark mass is far from understood, and in this paper we will report on a  study of this evolution in the kaon sector.

We compute the elastic scattering amplitudes for $\pi K$ in isospin-$\frac{1}{2}$ in $S$ and $P$ partial waves, using four values of the light-quark mass resulting in pion masses of approximately 239, 284, 329 \& 391 MeV. We find clear evidence for the vector $K^\star$ state for all values of the quark masses, while the $S$-wave appears qualitatively similar to experiment with a broad enhancement seen across the elastic region.

\emph{Methods} --- We utilize lattice Quantum Chromodynamics (QCD) as the only first-principles, systematically-improvable and generally-applicable approach to QCD. The use of a discretized Euclidean spacetime of finite volume allows us to determine hadronic correlation functions via Monte-Carlo sampling of gauge-fields. The Euclidean time-dependence of these correlation functions is controlled by the discrete spectrum of eigenstates of QCD in the finite volume.

This spectrum can be used to constrain the infinite-volume scattering amplitudes via the \emph{L\"uscher} method 
~\cite{Luscher:1990ux, Rummukainen:1995vs, Bedaque:2004kc, Kim:2005gf, Fu:2011xz, Leskovec:2012gb, Gockeler:2012yj,  He:2005ey, Lage:2009zv, Bernard:2010fp, Doring:2011vk, Doring:2011nd, Agadjanov:2013kja, Doring:2012eu, Hansen:2012tf, Briceno:2012yi, Guo:2012hv} -- for a recent review see Ref.~\cite{Briceno:2017max}. Through use of multiple lattice volumes and consideration of frames moving with respect to the lattice, sufficiently many energy levels can be obtained to determine in detail the energy-dependence of scattering amplitude across a large energy region. Several previous studies have considered $\pi K$ scattering using lattice QCD~\cite{Beane:2006gj, Fu:2011wc, Fu:2012tj, Lang:2012sv, Prelovsek:2013ela, Sasaki:2013vxa, Dudek:2014qha, Wilson:2014cna, Bali:2015gji, Brett:2018jqw, Helmes:2018nug}. 

The cubic nature of the periodic spatial boundary of the lattice means that states are characterized by 
irreducible representations (\emph{irreps}) of the cubic group, and of the relevant \emph{little groups} when considering moving frames. 
The mismatch between the cubic symmetry and the continuous rotational symmetry of the infinite-volume theory means that the irreps contain an infinite number of mixed partial waves. Because near threshold only a relatively small number of low partial waves are expected to be significant, in practice only two or three amplitudes influence the spectrum in the energy region we will consider.

To extract scattering amplitudes from finite-volume spectra it is important to accurately obtain all of the energy levels in the region of interest, and in order to do this, we compute matrices of correlation functions using a basis of operators, and diagonalize to obtain several excited energy eigenstates~\cite{Michael:1985ne,Blossier:2009kd}. A range of operators are considered, matching expectations of the kinds of finite-volume eigenstates in this case, consisting of $\bar{\psi}\Gamma {D}...{D} \psi$ constructions which resemble $q\bar{q}$ structures~\cite{Dudek:2010wm,Thomas:2011rh}, and meson-meson-like constructions~\cite{Dudek:2012gj}. The meson-meson operators are built from products of variationally-optimized meson operators, themselves sums of many $\bar{\psi}\Gamma D...D \psi$ constructions with the flavor and spin-parity of the relevant hadron: $\pi$, $K$ or $\eta$ in this instance. The virtue of this method is that excited state contaminations from the single-meson object contained within the meson-meson object are greatly reduced, and signals may then be obtained at earlier Euclidean times where statistical noise is typically lower.

\begin{table*}[bht]
\begin{tabular}{rlccc|lllll|c}
 
  $(L/a_s)^3$ & $\times \; T/a_t$ & $N_{\mathrm{cfgs}}$ & $N_\mathrm{vecs}$ & $N_\mathrm{t_{src}}$  
              & \multicolumn{1}{c}{$a_t m_\pi$} & \multicolumn{1}{c}{$a_t m_K$} & \multicolumn{1}{c}{$a_t m_\eta$} 
              & \multicolumn{1}{c}{$a_t m_\Omega$} & \multicolumn{1}{c|}{$\xi$} & $m_\pi$/MeV\\[0.2ex]
\hline
 {\scriptsize\{$16^3$, $20^3$, $24^3$\}} & $\times$ 128    &  {\scriptsize \{479,603,553\} }           & {\scriptsize \{64,128,162\}} 
                                  & 2--8  & 0.06906(13) & 0.09698(9)  & 0.10364(19)          & 0.2951(22) & 3.444(6)     & 391  \\ 
 $24^3$ &$\times$ 256 & 309 & 162 & 4--8  & 0.05593(28) & 0.09027(15) & 0.09790(100)         & 0.2857(8)  & 3.456(9)     & 327  \\
 $24^3$ &$\times$ 256 & 400 & 162 & 4     & 0.04735(22) & 0.08659(14) & 0.09602(70)          & 0.2793(8)  & 3.455(6)     & 284  \\
 $32^3$ &$\times$ 256 & 485 & 384 & 2--4  & 0.03928(18) & 0.08344(7)  & 0.09299(56)          & 0.2751(6)  & 3.453(6)     & 239  
\end{tabular}
\caption{
A summary of the lattices used in this study, with spatial volume $L^3$, temporal extent $T$, and the masses of relevant stable hadrons. 
$N_{\text{cfgs}}$ denotes the number of gauge configurations used and $N_{\text{vecs}}$ is the number of distillation vectors~\cite{Peardon:2009gh}. $N_\mathrm{t_{src}}$ is the number of different timeslices used for source operators. Brackets denote the uncertainty on the final digit.
}
\label{tab:lat}
\end{table*}

We make use of the \emph{distillation} method~\cite{Peardon:2009gh} that allows all of the Wick contractions specified by QCD to be efficiently obtained. Anisotropic lattices, having a finer spacing in time ($a_t$) than space ($a_s = a_t  \xi$), are used~\cite{Edwards:2008ja,Lin:2008pr}. Table~\ref{tab:lat} provides some details of these lattices -- the heaviest and lightest pion-mass lattices have been used previously to study many other channels~\cite{Dudek:2012gj, Dudek:2012xn, Briceno:2015dca, Briceno:2016kkp, Briceno:2016mjc, Dudek:2016cru, Moir:2016srx, Woss:2016tys, Briceno:2017qmb, Woss:2018irj,Wilson:2015dqa,Cheung:2016bym}, while the two intermediate pion-mass lattices are being used for the first time in this calculation. We have previously reported on $\pi K$ scattering on the 391 MeV lattice in refs.~\cite{Dudek:2014qha, Wilson:2014cna}, and we make use of these same spectra again. 

To quote results in physical units, the $\Omega$-baryon mass is used to set the scale via ${a_t^{-1}=\frac{m_\Omega^{\mathrm{phys.}}}{a_t m_\Omega^{\mathrm{latt.}}} }$, and in this paper, for all but the $m_\pi \approx 391$ MeV lattice, we use a new computation using 64 distillation vectors. For the $m_\pi \approx 239$ MeV lattice, this results in a more accurate value which supersedes that presented in Ref.~\cite{Wilson:2015dqa}. A more complete description of the methods used to arrive at the lattice QCD spectra is presented in Ref.~\cite{Wilson:2014cna}.

\begin{figure}
\includegraphics[width=1.0\columnwidth]{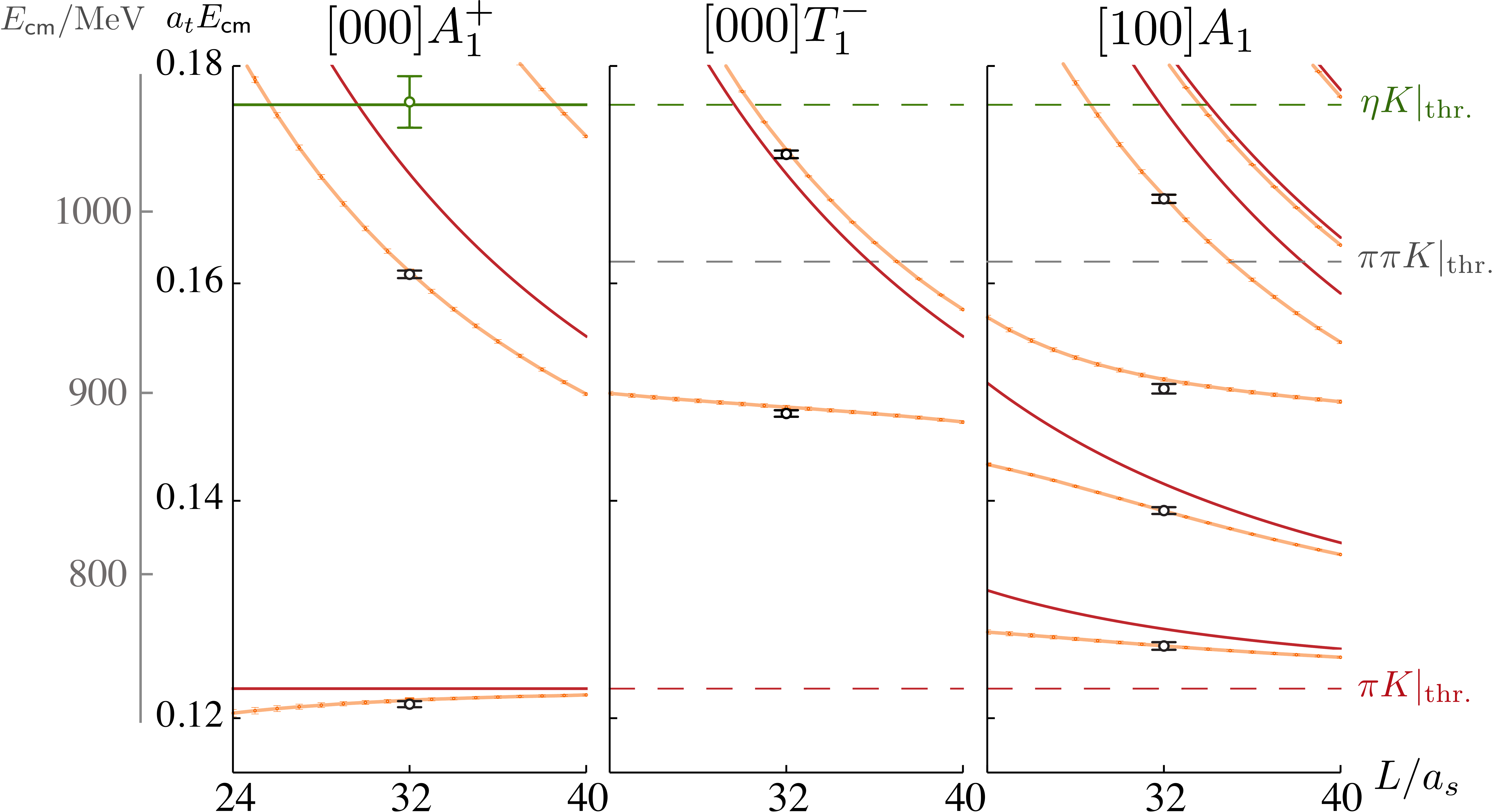}
\caption{An example of the finite-volume spectra computed with the ensemble corresponding to the smallest pion mass considered. The black points are finite-volume QCD energy levels used in obtaining the amplitudes. Green points indicate a level with only a significant contribution from an $\eta K$-like operator. Red and green curves indicate the positions of $\pi K$ and $\eta K$ energy levels in the absence of interactions, dashed lines indicate threshold energies.  The orange points and curves show the solutions of Eq.~\ref{eq_det} using a two-parameter $K$-matrix in $S$-wave and a Breit-Wigner in $P$-wave.}
\label{fig:spec_860}
\end{figure}

\emph{Finite-volume spectra} --- We show a representative sample of the spectra obtained in Fig.~\ref{fig:spec_860}, presenting two rest-frame spectra and one moving frame spectrum from the lightest pion mass considered.\footnote{Irreps are labeled as $[ijk]\Lambda^P$, where 3-momentum $\vec{k}=\frac{2\pi}{L}(i,j,k)=[ijk]$, $\Lambda$ is the irrep, and $P$ is the parity if $\vec{k}=\vec{0}$.}$^,$\footnote{We provide the remaining irreps and all of the spectra for the two new lattices, along with tables of operators used, in the supplemental material. Spectra on the heaviest pion-mass lattice are given in Ref.~\cite{Wilson:2015dqa}. In the region below $\eta K$ threshold, the qualitative pattern is the same across all 4 lattices.}$^,$\footnote{Spectra from the smallest mass lattice can be compared to those shown in Ref.~\cite{Brett:2018jqw}, which uses a subset of the same gauge configurations, the same quark mass parameters, but a different method for computing correlation functions.}
In $[000]A_1^+$, $S$-wave interactions dominate and we observe large shifts in energy away from expectations in a theory without $\pi K$ interactions -- there is an energy level below threshold, and another significantly below the next non-interacting energy, $\pi_{[100]} K_{[100]}$. At the $\eta K$ threshold a level appears that has significant overlap onto only the $\eta_{[000]} K_{[000]}$ operator, shown in green in Fig.~\ref{fig:spec_860}, and such a level persists across all the light-quark masses considered. 

The $[000]T_1^-$ irrep is dominated by $P$-wave with negligibly small contributions from $F$-wave and higher. An isolated level appears well below the lowest non-interacting energy, likely indicating the presence of a resonance which may be narrow given the relatively small shift of the next level up in energy. 

The denser spectrum in $[100]A_1$ reflects the contribution of both $S$ and $P$-wave amplitudes. This commonly occurs in unequal mass systems in moving frames since parity is not a good quantum number.

We can estimate the size of $D$-wave scattering in the elastic region, which can have an impact in many moving frame irreps, by considering the $[000]E^+$ irrep. Here a level is obtained at $a_t E_\mathsf{cm}=0.1699(3)$ on the smallest mass lattice, coincident with the expected non-interacting energy for $\pi_{[100]}K_{[100]}$. This energy corresponds to a negligibly small phase shift of $\delta_2^{\pi K} = 0.26(56)^\circ$, and similarly small values apply on the other lattices, such that we may neglect $D$-wave and higher partial-waves hereafter.

In our analysis, we choose to consider  only energies below the first inelastic threshold, which depending on the irrep and pion mass, is either $\pi\pi K$ or $\eta K$. This results in 28, 21, 18 and 36 energy levels to constrain $S$ and $P$-wave scattering amplitudes on lattices with pion masses of 239, 284, 329 and 391 MeV respectively.

\begin{figure}[tbh]
\includegraphics[width=1.0\columnwidth]{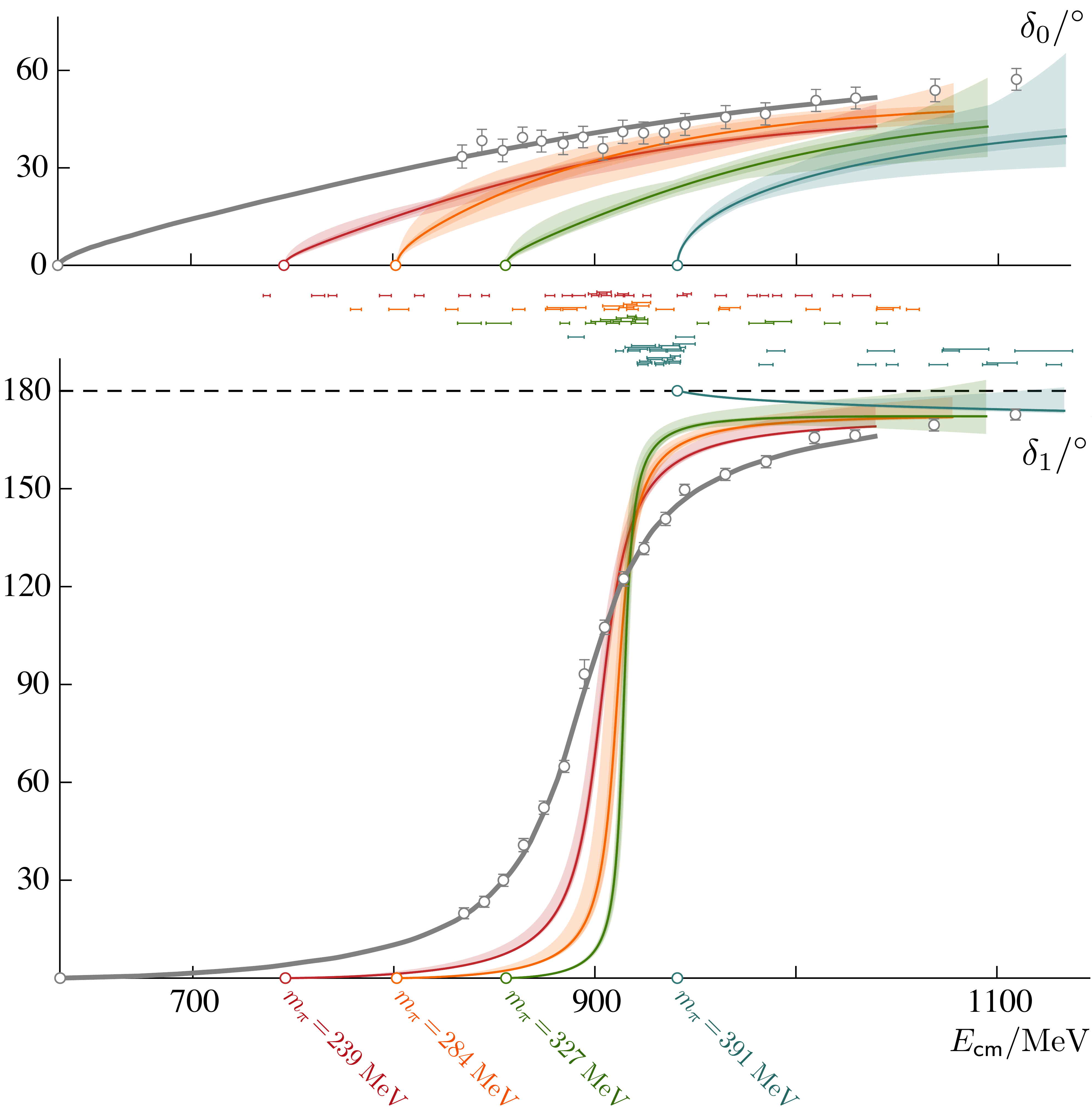}
\caption{$S$-wave (top) and $P$-wave (bottom) phase shifts. The central line and band correspond to the 2-parameter K-matrix in $S$-wave and a Breit-Wigner in $P$-wave described in the text and as used in Fig.~\ref{fig:spec_860}, the outer bands include the uncertainty over parameterizations, mass and anisotropy variations. The central coloured errorbars show the positions of the finite volume energy levels coloured by quark mass. The circles on the x-axes indicate $\pi K$ threshold at each mass. The grey points are experimental data from LASS~\cite{Estabrooks:1977xe, *Aston:1986jb, *Aston:1987ey, *Aston:1987ir} and the grey curves are from the phenomenological UFD parameterization of Ref.~\cite{Pelaez:2016tgi}.}
\label{fig:phases}
\end{figure}

\emph{Analysis} --- The relationship between the discrete spectrum in a finite volume, $\{E_n(L)\}$, and the infinite-volume scattering matrix, $\bm t(E)$, is given by the solutions of L\"{u}scher's determinant condition~\cite{Luscher:1990ux, Rummukainen:1995vs, Bedaque:2004kc, Kim:2005gf, Fu:2011xz, Leskovec:2012gb, Gockeler:2012yj,  He:2005ey, Lage:2009zv, Bernard:2010fp,  Agadjanov:2013kja, Doring:2012eu, Hansen:2012tf, Briceno:2012yi, Guo:2012hv, Briceno:2014oea}, 
\begin{align}
\det\Bigl[\bm{1}+i \rho(E)\, \bm{t}(E)\cdot\bigl(\bm{1}+i\bm{\mathcal{M}}(E,L)\bigr)\Bigr]=0,
\label{eq_det}
\end{align}
where $\bm{\mathcal{M}}(E,L)$ is a matrix of known functions in the space of partial-waves for each irrep,\footnote{This only works for energies below three-particle thresholds. For on-going efforts to remove this restriction, see Refs.~\cite{Briceno:2018aml, Briceno:2017tce,Hansen:2014eka, Hansen:2019nir}.} and $\rho=\frac{2k_\mathsf{cm}}{E}$. To overcome the dependence of each energy level on $t(E)$ for multiple partial-waves, we parameterize the energy dependence of the scattering amplitudes using a variety of forms which respect unitarity and which have sufficient freedom to describe the spectra. For a given parameterization, the parameter values are found which upon solving the above determinant equation give finite-volume spectra that best describe the lattice spectra\footnote{minimizing the correlated $\chi^2$ presented in Ref.~\cite{Wilson:2014cna}}. A representative example in which $S$-wave and $P$-wave are parameterized is shown by the orange points and curves in Fig.~\ref{fig:spec_860}.

To avoid bias, we consider a wide selection of scattering amplitude parameterizations that fall into four familiar categories: effective-range expansions, Breit-Wigners, $K$-matrices, as given in Ref.~\cite{Wilson:2014cna} in Eqs.~9-13, and unitarized chiral perturbation theory (U$\chi$PT)~\cite{Truong:1988zp,Dobado:1989qm,GomezNicola:2001as,GomezNicola:2007qj}. The $K$-matrix features the most flexibility, and we opt to use the Chew-Mandelstam phase-space in which a logarithm is generated from the known imaginary part from unitarity~\cite{Wilson:2014cna}. Our $K$-matrix forms respect $s$-channel unitarity, but do not include any features from scattering in the cross-channels (no ``left-hand cuts''). 

The U$\chi$PT amplitudes share the logarithm mentioned above associated with the $s$-channel cut, but they also contain perturbative features associated with the cross-channels. All the masses considered here are far from the chiral SU(3) symmetric point about which these amplitudes are expanded. The amplitudes would break unitarity without a unitarization step which, although not unique, results in U$\chi$PT amplitudes that respect unitarity perturbatively. We choose to apply the $\mathcal{O}(p^4)$ SU(3) amplitudes specifically because they have been used in studying the pion mass dependence of $\pi K$ scattering in Ref.~\cite{Nebreda:2010wv}. 


The $S$ and $P$-wave phase-shifts of all considered amplitude parameterizations which can describe the finite-volume spectra with $\chi^2/N_\mathrm{dof}$ below $2.0$ are plotted in Fig.~\ref{fig:phases} -- there are 14-17 per pion mass, and a complete list can be found in the supplemental material. The central curves are from a four parameter fit with a Breit-Wigner in $P$-wave and a two-parameter $K$-matrix, linear in $s=E_\mathsf{cm}^2$, in $S$-wave. This same choice is used to produce the orange curves in Fig.~\ref{fig:spec_860}.
Very little variation is seen between parameterizations -- the amplitudes are well-determined and there is little sensitivity to the precise form used. As the pion mass reduces we see a clear trend towards the experimental phase-shifts. The striking difference in the $P$-wave amplitude between the $m_\pi \approx 391$ MeV and 327 MeV lattices is caused by the $K^\star$ changing from a bound state below $\pi K$ threshold to a resonant $K^\star$ above threshold.

\begin{figure}
\includegraphics[width=1.0\columnwidth]{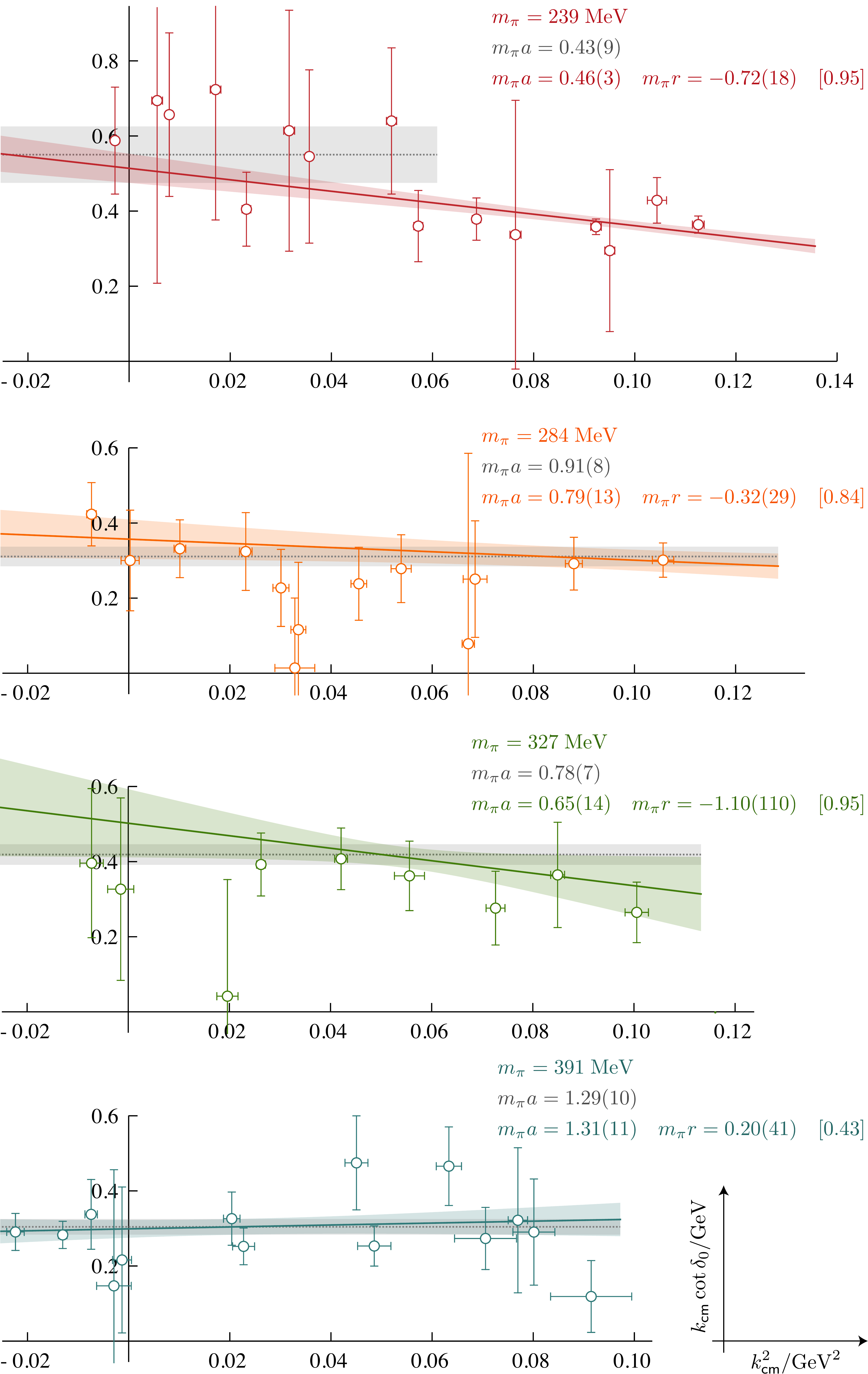}
\caption{The $S$-wave amplitudes shown as $k\cot\delta_0$. The discrete points are as described in the text.
Points in the region of the $K^\star$ pole appear particularly sensitive to the $P$-wave amplitude and provide little constraint in $S$-wave -- several have been removed from this plot. Curves correspond to the $S$-wave scattering length and effective range fits with a 3-parameter $K$-matrix in $P$-wave. 
Square brackets show parameter correlations.}
\label{fig:kcot_S}
\end{figure}

The $S$-wave amplitude is presented in a different manner in Figure 3, via $k_\mathsf{cm}\, \cot \delta_0$, the quantity which has an effective range expansion $\frac{1}{a} + \tfrac{1}{2} r k_\mathsf{cm}^2 + \ldots$, where $a$ is the scattering length and $r$ is the effective range.  The discrete points shown reflect the $S$-wave for a fixed $P$-wave three-parameter $K$-matrix amplitude, with the uncertainty including a sampling of several amplitudes, while the curves show scattering-length and effective-range amplitudes with the same three-parameter $K$-matrix in $P$-wave. It is clear that for the three largest pion masses, the amplitude over the whole elastic region is acceptably well described by just a scattering-length, while at the smallest mass an additional effective-range term is required. A clear trend of decreasing $m_\pi a$ is observed with decreasing light-quark mass, which is qualitatively consistent with leading order chiral perturbation theory. Fig.~\ref{fig:kcot_S} shows little evidence for a large effective range parameter that might signal the presence of a narrow resonance, nor for any strong enhancement below threshold that would be suggestive of important effects from an \emph{Adler zero} in $t(E)$~\cite{Adler:1964um,Adler:1965ga}.


\begin{figure}
\includegraphics[width=1.0\columnwidth]{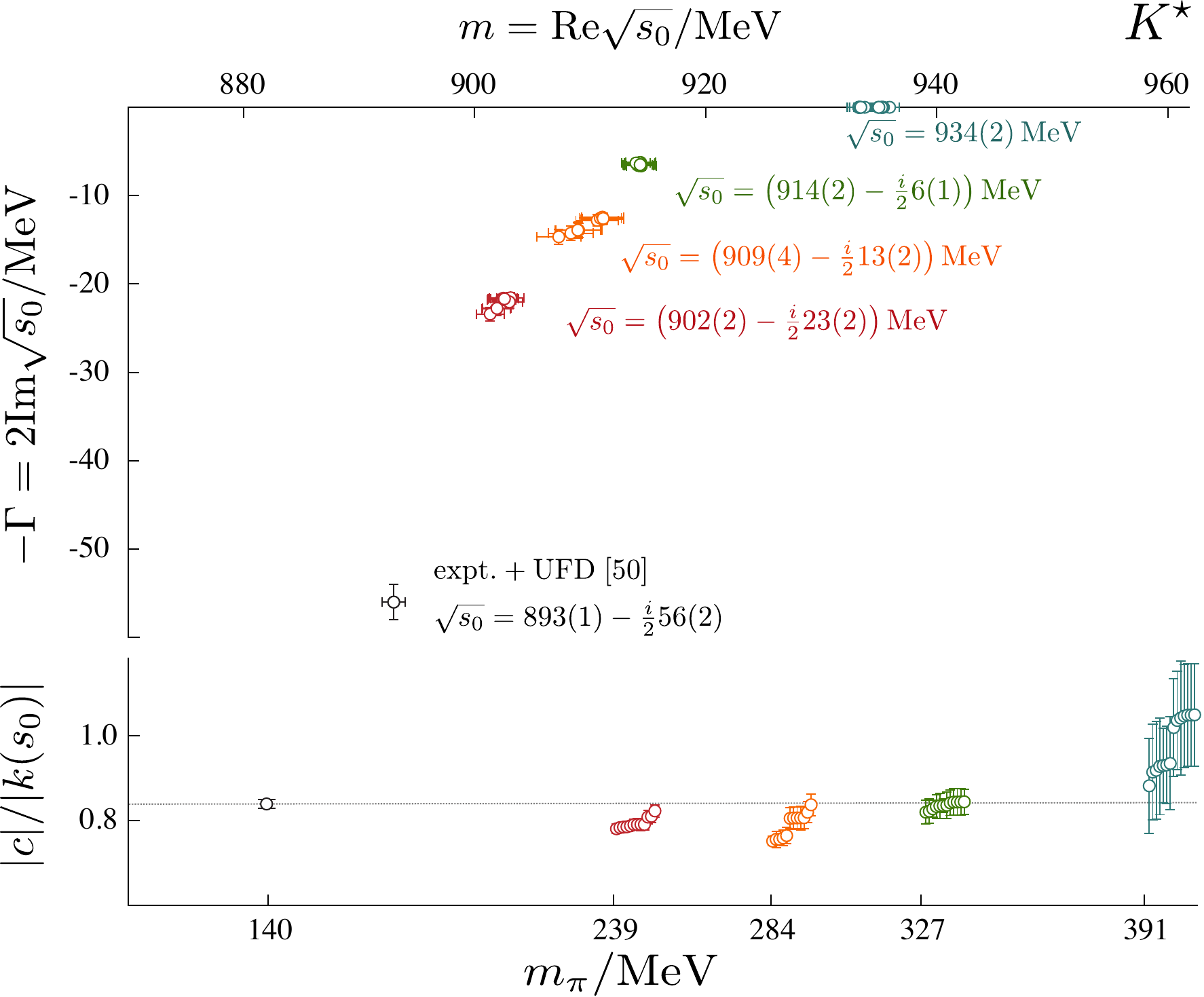}
\caption{Vector $K^\star$ pole positions (top) and couplings (bottom) across the four pion masses used here compared to a phenomenological fit to experimental data from Ref.~\cite{Pelaez:2016tgi} shown in black (also used in Fig.~\ref{fig:phases}). The pole positions quoted at each mass include statistical uncertainties and an uncertainty from sampling many parameterizations. For the largest pion mass, the $K^\star$ pole appears as a bound state, it otherwise appears as a complex pole on the unphysical sheet.}
\label{fig:poles_P}
\end{figure}

\emph{Poles} --- The singularity content of a scattering amplitude, considered as a function of complex energy, is closely connected to its spectroscopic content, with the presence of a \emph{pole}, $t\sim c^2/\left(s_0-s\right)$ on an \emph{unphysical} Riemann sheet,
typically interpreted as being the most rigorous signal for a resonance. The pole position can be related to the mass $m_R$ and width $\Gamma_R$, $\sqrt{s_0} = m_R \pm i \Gamma_R/2$, and the residue gives access to the \emph{coupling}, $c$. 

All the $P$-wave amplitudes that we found were able to describe our finite-volume spectra feature a single pole close to the real axis that we summarize in Fig.~\ref{fig:poles_P}. For the heaviest pion mass we considered, the pole is on the real axis, corresponding to a stable bound state, but otherwise it is off the real axis, corresponding to an unstable resonance. A smooth evolution is seen with an approximately flat effective coupling $|c|/|k|$ as a function of the pion mass. The scatter due to parameterization choice is observed to be quite modest, comparable to the size of the statistical uncertainty.

The $S$-wave amplitude presented in Figures~\ref{fig:phases} and \ref{fig:kcot_S} is superficially very simple: there is a rising phase shift, usually attributed to an attractive system, but no sharp features that signal the presence of a nearby pole or other singularity. This suggests that if any resonance pole is influencing this behavior, it must lie far into the complex plane.
In order to determine such distant poles, it is necessary to consider the features of partial-wave amplitudes at complex $s$ which arise due to known properties of scattering, like crossing symmetry and unitarity. In elastic $\pi K$ scattering, the complex plane contains three cuts~\cite{Mandelstam:1958xc, Oller:1998zr, *Cherry:2000ut, *Buettiker:2003pp, *Zheng:2003rw, *Zhou:2006wm, *DescotesGenon:2006uk, *Pelaez:2018qny}: in addition to the $s$-channel unitarity cut, which is correctly handled in the finite-volume formalism, unitarity in the cross-channels leads to a circular cut at $|s|=(m_K^2-m_\pi^2)$ and a left-hand cut that spans $-\infty<s<(m_K-m_\pi)^2$. If these cuts are as close to the elastic scattering region as any hypothetical resonance pole, their effect must be accounted for if the pole is to be accurately determined. Of the amplitudes applied here, only U$\chi$PT has any contributions from the cross-channels, and the degree to which they are correctly handled has been debated~\cite{Boglione:1996uz,Nieves:2001de,Cavalcante:2001yw,GomezNicola:2007qj}. When the U$\chi$PT amplitudes have their parameter freedom constrained by the finite-volume spectra presented above, a complex pole is found with a real energy around $m_\pi+m_K$ and a large imaginary part, not dissimilar to the experimental $\kappa$ resonance. In addition, many of the $K$-matrix forms we implement, which lack any explicit left-hand cut behavior, also feature poles at similar energies; however, some do not and many have other nearby poles. Even with precise information about the amplitude for real energies, the analytic continuation required to reach any pole is sufficiently large that a unique result is not found.

\vspace{1mm}

\emph{Summary} --- We have extracted $S$ and $P$-wave elastic isospin-$\frac{1}{2}$ $\pi K$ scattering amplitudes from lattice QCD spectra using L\"{u}scher's formalism at four light-quark masses, corresponding to pion masses between 239 and 391 MeV. The resulting amplitudes show a smooth evolution towards experimental data as the quark mass approaches its physical value. Continuing the $P$-wave amplitude to complex values of the energy, the lowest-lying vector $K^\star$ resonance appears as a pole singularity in a way consistent with the canonical picture of a compact quark-antiquark state that acquires a decay width by coupling to the meson-meson continuum. 
The $S$-wave amplitudes are well-determined for real energies, however the analytic continuation into the complex plane does not yield a unique result that we can interpret in terms of the $\kappa$ pole. Along with our previous study of the $\sigma$~\cite{Briceno:2016mjc}, this provides motivation for future analyses that incorporate now-standard lattice QCD analysis techniques, namely \emph{L\"uscher-like} analysis of finite-volume spectra, and those in use in the amplitude analysis community, e.g. \emph{Roy-Steiner equations}, which account for the known singularities due to cross-channel physics. In the current case, an input to such a calculation would be information about  $\pi\pi\to K\overline{K}$ in several partial waves, which can be obtained in a lattice calculation using the generalization of the L\"uscher formalism for coupled-channels~\cite{Briceno:2017qmb, Wilson:2015dqa}. 

In closing, we believe this poses a timely challenge for the lattice QCD and amplitude-analysis communities to address jointly. In so doing, we will not only be able to acquire a detailed picture of the mysterious $\sigma$ and $\kappa$ resonances, but an understanding of the breaking of SU(3) flavor symmetry and thus the origin and nature of these resonances.


\acknowledgments
{
  We thank our colleagues within the Hadron Spectrum Collaboration, with particular thanks to Antoni Woss for useful discussions. We also thank B\'alint Jo\'o for help and the use of the {\tt QPhiX} codes, and Kate Clark for use of the {\tt QUDA} codes.
	DJW acknowledges support from a Royal Society--Science Foundation Ireland University Research Fellowship Award UF160419.
	RAB, JJD and RGE acknowledge support from the U.S. Department of Energy contract DE-AC05-06OR23177, under which Jefferson Science Associates, LLC, manages and operates Jefferson Lab. 
        RAB also acknowledges support from the U.S. Department of Energy Early Career award contract DE-SC0019229.
	JJD also acknowledges support from the U.S. Department of Energy award contract DE-SC0018416.
        CET acknowledges support from the U.K. Science and Technology Facilities Council (STFC) [grant number ST/P000681/1].

The software codes
{\tt Chroma}~\cite{Edwards:2004sx}, {\tt QUDA}~\cite{Clark:2009wm,Babich:2010mu}, {\tt QUDA-MG}~\cite{Clark:SC2016}, {\tt QPhiX}~\cite{ISC13Phi}, and {\tt QOPQDP}~\cite{Osborn:2010mb,Babich:2010qb} were used. 
The authors acknowledge support from the U.S. Department of Energy, Office of Science, Office of Advanced Scientific Computing Research and Office of Nuclear Physics, Scientific Discovery through Advanced Computing (SciDAC) program.
Also acknowledged is support from the U.S. Department of Energy Exascale Computing Project.
The contractions were performed on clusters at Jefferson Lab under the USQCD Initiative and the LQCD ARRA project. This research was supported in part under an ALCC award, and used resources of the Oak Ridge Leadership Computing Facility at the Oak Ridge National Laboratory, which is supported by the Office of Science of the U.S. Department of Energy under Contract No. DE-AC05-00OR22725.
This research is also part of the Blue Waters sustained-petascale computing project, which is supported by the National Science Foundation (awards OCI-0725070 and ACI-1238993) and the state of Illinois. Blue Waters is a joint effort of the University of Illinois at Urbana-Champaign and its National Center for Supercomputing Applications. This research used resources of the National Energy Research Scientific Computing Center (NERSC), a DOE Office of Science User Facility supported by the Office of Science of the U.S. Department of Energy under Contract No. DE-AC02-05CH11231.
The authors acknowledge the Texas Advanced Computing Center (TACC) at The University of Texas at Austin for providing HPC resources. Gauge configurations were generated using resources awarded from the U.S. Department of Energy INCITE program at Oak Ridge National Lab, and also resources awarded at NERSC. 
}


\bibliographystyle{apsrev4-1}
\bibliography{bib.bib}

\clearpage

\appendix

\onecolumngrid
\section*{Supplemental material}


\subsection{Spectra}

\begin{figure*}[!ht]
\includegraphics[width=1.0\textwidth]{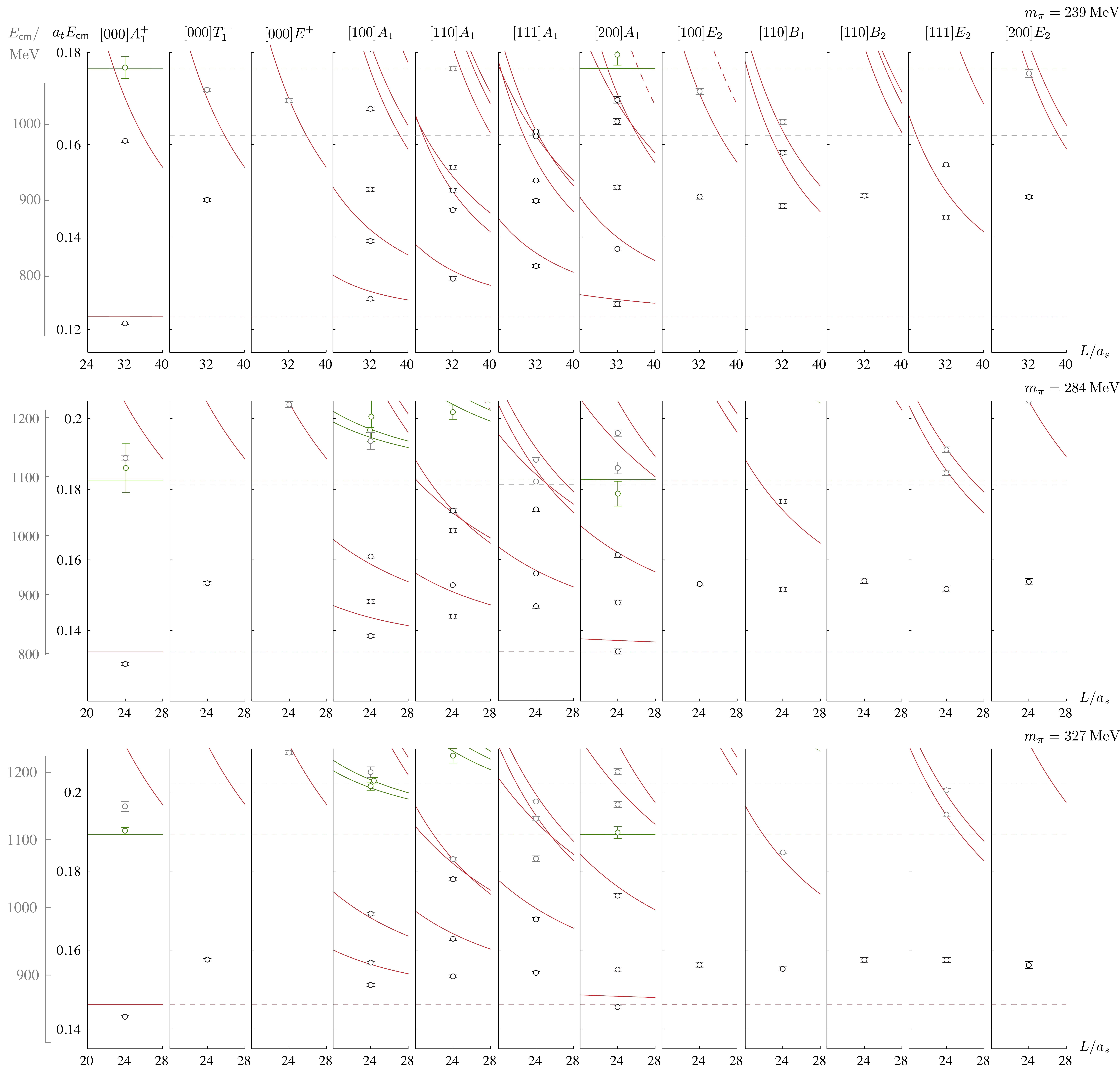}
\caption{Finite-volume spectra from the smallest three light-quark mass lattices considered in this study. The black points show energy levels that are used to obtain the scattering amplitudes; grey and green points are energy levels that have not been used to obtain the scattering amplitudes. Solid curves denote non-interacting energies. Dashed coloured lines denote $\pi K$ threshold in red, and $\eta K$ threshold in green. Dashed grey lines show the position of $\pi\pi K$ threshold. The operators used to obtain these energy levels are given in Table~\ref{tab:ops}.}
\label{fig_all_spec}
\end{figure*}

The spectra obtained in this study are shown in Fig~\ref{fig_all_spec}. 

\clearpage

\subsection{Highlighted parameterizations}

In the text we give three pairs of parameterizations special attention -- a ``reference'' parameterization that features in Figs.~\ref{fig:spec_860} and~\ref{fig:phases} is given by a linear polynomial in $s=E_\mathsf{cm}^2$,
\begin{align}
t^{-1}(s)=K^{-1}(s)+I(s),
\quad
K(s) &=\gamma_0 + \gamma_1\hat s, 
\quad 
\hat{s}=\tfrac{s-s_{\mathrm{thr.}}}{s_\mathrm{thr.}},
\label{eq_KmatS}
\end{align}
where $\gamma_i$ are free parameters and $s_\mathrm{thr.}=(m_\pi+m_K)^2$ for the $S$-wave amplitude, and a Breit-Wigner in $P$-wave given by Eq.~10 from Ref.~\cite{Wilson:2014cna}. The Chew-Mandelstam phase space $I$ from appendix B of Ref.~\cite{Wilson:2014cna} is used in $S$-wave with the subtraction point fixed at $\pi K$ threshold. This has a logarithmic real part for real $s$ above threshold, and better properties when analytically continued compared to using the simpler phase space $I=-i\rho=-i2k_\mathsf{cm}/\sqrt{s}$, where 
\begin{align}
k_\mathsf{cm}(s)= \frac{1}{2s^{1/2}}\left(s-(m_\pi+m_K)^2\right)^{1/2}\left(s-(m_K-m_\pi)^2\right)^{1/2} \, .
\end{align}
The numerical results of using this pair of parameterizations across the four light-quark masses are given in Table~\ref{tab:ref_fits}.

\begin{table}[bh]
\begin{tabular}{c|c|ccc|ccc}
$m_\pi/$MeV   & $\chi^2/{N_{\mathrm{dof}}}$ &  $\gamma_0$ &  $\gamma_1$        & corr.  & $m_R$/MeV & $g_R$ & corr. \\
\hline\hline
\multirow{1}{*}{239}           
              &   $\frac{39.15}{28-4}=1.63$  &  $0.48(1)$   & $2.29(19)$     & -0.57 & 904(1) & 4.81(7)  & 0.18   \\[0.5ex] 
\multirow{1}{*}{284}       
              & $\frac{20.13}{21-4}=1.18$    &  $0.90(15)$   & $3.34(64)$     & -0.33 & 911(2) & 4.67(11) &  0.01   \\[0.5ex]          
\multirow{1}{*}{327}
              & $\frac{25.31}{18-4}=1.81$    &  $0.59(16)$   & $3.14(90)$     & -0.75 & 915(2) & 5.20(16) &  0.28   \\[0.5ex] 
\multirow{1}{*}{391}           
              & $\frac{51.36}{36-4}=1.60$    &  $0.77(24)$   & $1.20(39)$     & -0.91 & 936(1) & 5.52(20) & -0.31
\end{tabular}
\caption{Reference amplitude fits corresponding to a linear-order $K$-matrix with parameters $\gamma_0$, $\gamma_1$ in $S$-wave, and a Breit-Wigner in $P$-wave with a Breit-Wigner mass parameter $m_R$ and Breit-Wigner coupling $g_R$.\vspace{0.2cm}}
\label{tab:ref_fits}
\end{table}

The other two pairs of parameterizations highlighted are scattering lengths and effective ranges in $S$-wave, defined by $k_\mathsf{cm}^{2\ell+1}\cot\delta_\ell=\frac{1}{a_\ell}+\frac{1}{2}r_\ell k^2_\mathsf{cm}+ \mathcal{O}(k^4_\mathsf{cm})$, with $\ell=0$, and a three-parameter ``pole plus constant'' $K$-matrix parameterization in $P$-wave,
\begin{align}
t^{-1}(s)=(2 k_\mathsf{cm}(s)\cdot K(s) \cdot 2 k_\mathsf{cm}(s))^{-1}+I(s),
\quad
K=\frac{g^2}{m^2-s}+\gamma_0\,,
\label{eq_KP}
\end{align}
with a Chew-Mandelstam phase space $I(s)$, which is a specific case of Eq.~11 and Eq.~12 in Ref.~\cite{Wilson:2014cna}. We make use of these two pairs of amplitudes in producing Fig.~\ref{fig:kcot_S}.
The effective range $S$-wave amplitudes do not have the logarithmic term of the Chew-Mandelstam phase space $I(s)$ and thus are relatively badly behaved deep in the complex plane.

\begin{table}[bh]
\begin{tabular}{c|c|ccc|ccc}
$m_\pi/$MeV   & $\chi^2/{N_{\mathrm{dof}}}$ &  $m_\pi a$ &  $m_\pi r$           & corr.  & $m$/MeV & $g$ & $\gamma_0$/GeV$^{-2}$ \\
\hline\hline
\multirow{2}{*}{239}           
              &     $\frac{10.84}{15-4}=0.99$  &     $0.43(9)$   & $-$         & $-$  & 901(2) & 0.367(39) & 0.88(39) \\[0.5ex] 
              &     $\frac{36.58}{28-5}=1.59$  &     $0.46(3)$   & $-0.72(18)$ & 0.95 & 902(1) & 0.403(20) & 0.29(9)\\[0.5ex] 
\hline
\multirow{2}{*}{284}       
              & $\frac{18.79}{21-4}=1.11$      &     $0.91(8)$   & $-$         & $-$  & 908(2) & 0.410(11) & 0.30(8)\\[0.5ex]
              & $\frac{17.27}{21-5}=1.08$      &     $0.79(13)$    & $-0.32(29)$ & 0.84 & 908(2) & 0.411(12) & 0.29(8)\\[0.5ex] 
\hline
\multirow{2}{*}{327}
              & $\frac{25.86}{18-4}=1.85$      &  $0.78(7)$  & $-$              & $-$  & 914(2) & 0.412(13) & 0.31(13)  \\[0.5ex] 
              & $\frac{24.70}{18-5}=1.90$      &  $0.65(14)$   & $-1.10(110)$     & 0.95 & 915(2) & 0.412(14) & 0.08(25)  \\[0.5ex] 
\hline
\multirow{2}{*}{391}           
              & $\frac{31.90}{36-4}=1.00$      &  $1.29(10)$   & $-$            & $-$  & 933(1) & 0.522(21) & 0.55(13)   \\[0.5ex] 
              & $\frac{31.65}{36-5}=1.02$      &  $1.31(11)$   & $0.20(41)$     & 0.43 & 933(1) & 0.521(21) & 0.56(15)
\end{tabular}
\caption{Scattering length and effective range parameters obtained from fitting the finite-volume spectra, corresponding to the bands in Fig.~\ref{fig:kcot_S}. A three-parameter $K$-matrix, as described in the text, has been fitted simultaneously in $P$-wave.}
\label{tab:er_fits}
\end{table}


\clearpage

\subsection{Explicit list of parameterizations}

The amplitudes described above are used, and additionally the following,
\begin{align}
K(s)^{-1}&=\sum_{i=0}^{N_S} c_i \hat{s}^i\label{eq_Kinv}\\
K(s)&=\left(\sum_{i=0}^{N_{S_\gamma}} \gamma_i \hat{s}^i\right)/\left(1+\sum_{i=1}^{N_{S_c}} c_i \hat{s}^i\right)
\label{eq_Kinv_rat}\\
K(s)^{-1}&=\frac{1}{s-s_A}\left(\sum_{i=0}^{N_{S_c}} c_i \hat{s}^i\right)
\label{eq_Kinv_sA}
\end{align}
where $\gamma_i$, $c_i$ are free parameters. In Eq.~\ref{eq_Kinv_sA}, which is just a special case of Eq.~\ref{eq_Kinv_rat}, $s_A$ is fixed at the Adler zero position from leading order chiral perturbation theory. We also use the effective range expansion as described above. In Table~\ref{tab:all_amps} we refer to the leading order effective range $1/a_\ell$ as SL and the linear order effective range version as ER. 

In $P$-wave we make use of a Breit-Wigner parameterization and $K$-matrix parameterizations,
\begin{align}
K(s)&=\frac{\left(\sum_{i=0}^{N_g} g_i s^i\right)^2}{m^2-s}+\sum_{i=0}^{N_P} \gamma_i \hat{s}^i\,.
\label{eq_Kmat_grun}
\end{align}
We summarize the results of fitting spectra using these amplitudes in Table~\ref{tab:all_amps}. Several of the amplitudes used did not perform well at every value of light-quark masses -- these are marked in italics. Some amplitudes also produced unphysical features such as nearby physical sheet singularities -- several of these were also excluded and are marked in italics. Many more parameterisations were tested, we have chosen to show only parameterisations which were successful on at least one of the lattices.

The columns $I_S$ and $I_P$ indicate the phase spaces used -- CM corresponds to Chew-Mandelstam, $-i\rho$ is the simple phase described above, BW indicates the Breit-Wigner formula was used. Up to arbitrary subtractions, the U$\chi$PT amplitudes have the same logarithm as the Chew-Mandelstam phase space, indicated by $\sim$CM.

\newcommand{\rKg}[0]{Eq.~\ref{eq_Kmat_grun}}
\newcommand{\rKS}[0]{Eq.~\ref{eq_KmatS}}
\newcommand{\rBW}[0]{BW}
\newcommand{\rKA}[0]{Eq.~\ref{eq_Kmat_sA}}
\newcommand{\rKP}[0]{Eq.~\ref{eq_KP}}
\newcommand{\rKi}[0]{Eq.~\ref{eq_Kinv}}
\newcommand{\rKj}[0]{Eq.~\ref{eq_Kinv_sA}}	
\newcommand{\rSL}[0]{SL}
\newcommand{\rER}[0]{ER}
\newcommand{\rKr}[0]{Eq.~\ref{eq_Kinv_rat}}
\newcommand{\np}[0]{$N_\mathrm{pars.}$}
\newcommand{\ir}[0]{-$i\rho$}
\newcommand{\ik}[0]{-$i k_\mathsf{CM}$}

\begin{table*}[bht]
  \begin{tabular}{cc|ccc|cc|c|ccccccc} 
        $S$-amp &$P$-amp & $N_S$& $N_P$ &$\quad N_g \quad$& $I_S$& $I_P$& \np         & \multicolumn{4}{c}{$\chi^2/N_{\mathrm{dof}}$ for $m_\pi/\mathrm{MeV}$} & ~ \\
                      &        &      &       &      &      &          & $S$+$P$     & $\approx$ 391 & $\approx$ 327 & $\approx$ 286 & $\approx$ 239 \\
       
       \hline
 \rSL   & \rBW    & -    &  -   &  - & \ir  & BW    & 1+2        & $\frac{49.76}{36-3}=1.51$ & $\it \frac{32.87}{18-3}=2.19$    & $\it \frac{38.73}{21-3}=2.15$  & $\it \frac{65.63}{28-3}=2.63$   \\
 \rER   & \rBW    & -    &  -   &  - & \ir  & BW    & 2+2        & $\frac{49.72}{36-4}=1.55$ & $\frac{24.79}{18-4}=1.77$        & $\frac{33.50}{21-4}=1.97$      & $\frac{45.98}{28-4}=1.92$   \\
 \rSL   & \rKP    & -     &  -  &  - & \ir  & CM    & 1+3        & $\frac{31.90}{36-4}=1.00$ & $\frac{25.86}{18-4}=1.85$        & $\frac{18.79}{21-4}=1.11$      & $\it \frac{59.92}{28-4}=2.50$ \\
 \rER   & \rKP    & -     &  -  &  - & \ir  & CM    & 2+3        & $\frac{31.65}{36-5}=1.02$ & $\frac{24.70}{18-5}=1.90$        & $\frac{17.27}{21-5}=1.08$      & $\frac{36.58}{28-5}=1.59$ \\
    \hline
 \rKS   & \rBW   & 0    &  -    &  - & CM   & BW    & 1+2        & $\frac{62.75}{36-3}=1.90$ & $\it \frac{59.08}{18-3}=3.94$    & $\it \frac{108.3}{21-3}=6.01$  & $\it \frac{241.47}{28-3}=9.66$ \\
 \rKS   & \rBW   & 1    &  -    &  - & CM   & BW    & 2+2        & $\frac{51.36}{36-4}=1.60$ & $\frac{25.31}{18-4}=1.81$        & $\frac{20.13}{21-4}=1.18$      & $\frac{39.15}{28-4}=1.63$ \\
 \rKS   & \rKP   & 0    &  -    &  - & CM   & CM    & 1+3        & $\frac{42.21}{36-4}=1.32$ & $\it \frac{31.20}{18-4}=2.23$    & $\it \frac{70.80}{21-4}=4.16$  & $\it \frac{239.23}{28-4}=9.97$\\
 \rKS   & \rKP   & 1    &  -    &  - & CM   & CM    & 2+3        & $\frac{31.93}{36-5}=1.03$ & $\frac{24.54}{18-5}=1.89$        & $\frac{12.50}{21-5}=0.78$      & $\frac{37.70}{28-5}=1.64$\\
 \rKS   & \rKg   & 1    &  1    &  0 & CM   & CM    & 2+4        & $\it \frac{28.74}{36-6}=0.96$ & $\frac{23.09}{18-6}=1.92$    & $\frac{12.48}{21-6}=0.83$      & $\frac{37.70}{28-6}=1.71$\\
 \rKS   & \rKg   & 1    &  -    &  1 & CM   & CM    & 2+3        & $\frac{33.00}{36-5}=1.06$ & $\frac{24.64}{18-5}=1.90$        & $\frac{12.48}{21-5}=0.78$      & $\frac{37.70}{28-5}=1.64$ \\
 \rKS   & \rKg   & 1    &  0    &  1 & CM   & CM    & 2+4        & $\frac{31.93}{36-6}=1.06$ & $\it \frac{24.54}{18-6}=2.04$    & $\frac{12.48}{21-6}=0.83$      & $\frac{37.70}{28-6}=1.71$ \\
    \hline
 \rKi   & \rBW   & 1    &  -    &  - & CM   & BW    & 2+2        & $\it \frac{47.88}{36-4}=1.50$ & $\frac{25.20}{18-4}=1.80$    & $\it \frac{39.83}{21-4}=2.34$  & $\it \frac{53.67}{28-4}=2.24$ \\
 \rKi   & \rKP   & 1    &  -    &  - & CM   & CM    & 2+3        & $\frac{31.20}{36-5}=1.01$ & $\frac{25.09}{18-5}=1.93$        & $\frac{20.94}{21-5}=1.31$      & $\frac{42.40}{28-5}=1.84$ \\
    \hline
 \rKr  & \rBW & 1$^{(c)}$,1$^{(\gamma)}$ & - & - & CM & BW & 3+2 & $\frac{40.76}{36-5}=1.31$ & $\frac{24.44}{18-5}=1.88$         & $\frac{19.89}{21-5}=1.24$     & $\frac{37.01}{28-5}=1.61$   \\
 \rKr  & \rKP & 1$^{(c)}$,1$^{(\gamma)}$ & - & - & CM & CM & 3+3 & $\frac{28.27}{36-6}=0.94$ & $\it \frac{24.34}{18-6}=2.03$     & $\frac{12.49}{21-6}=0.83$     & $\frac{32.32}{28-6}=1.47$   \\
    \hline
 \rKj   & \rBW   & 1    &  -    &  - & CM   & BW    & 2+2        & $\frac{51.30}{36-4}=1.60$ & $\frac{24.48}{18-4}=1.75$        & $\frac{25.89}{21-4}=1.52$      & $\frac{38.60}{28-4}=1.61$ \\
 \rKj   & \rKP   & 1    &  -    &  - & CM   & CM    & 2+3        & $\frac{31.93}{36-5}=1.03$ & $\frac{24.39}{18-5}=1.88$        & $\frac{13.73}{21-5}=0.86$      & $\frac{32.41}{28-5}=1.41$ \\
    \hline
 \multicolumn{2}{c|}{  U$\chi$PT  }      
                         & -    &  -  &  -  
                                        & \multicolumn{2}{c|}{$\sim$CM}
                                                    & 4          & $\frac{38.18}{31-4}= 1.41$ & $\frac{26.06}{18-4}=1.86$        & $\frac{25.88}{21-4}=1.52$     & $\frac{39.32}{28-4}= 1.64$  \\
 \multicolumn{2}{c|}{  U$\chi$PT  }      
                         & -    &  -  &  -  
                                        & \multicolumn{2}{c|}{$\sim$CM}
                                                    & 4          & $\frac{49.29}{31-4}= 1.64$ & $\frac{24.41}{18-4}=1.74$        & $\frac{25.26}{21-4}=1.49$     & $\frac{39.66}{28-4}= 1.65$  
  \end{tabular}

\caption{The parameterizations used in this study. $\chi^2$ values in italics denote an amplitude that was rejected due to either a too large $\chi^2/N_\mathrm{dof}$ or nearby singularities not supported by the finite volume spectra.}
\label{tab:all_amps}
\end{table*}

\clearpage
\subsection{Operator lists}
We use the operators listed in Table~\ref{tab:ops} in obtaining the energy levels shown in Fig.~\ref{fig_all_spec}.
\begin{table}[htb]
\begin{tabular}{l|c|c|c}
                           & \multicolumn{1}{c|}{$m_\pi\approx239$ MeV}              & \multicolumn{1}{c|}{$m_\pi\approx286$ MeV}              & \multicolumn{1}{c}{$m_\pi\approx327$ MeV} \\
                           & \multicolumn{1}{c|}{$a_t m_\ell=-0.0860$, $a_t m_s=-0.0743$} & \multicolumn{1}{c|}{$a_t m_\ell=-0.0856$, $a_t m_s=-0.0743$} & \multicolumn{1}{c}{$a_t m_\ell=-0.0850$, $a_t m_s=-0.0743$} \\
                           & \multicolumn{1}{c|}{$L/a_s=32$, $T/a_t=256$}           & \multicolumn{1}{c|}{$L/a_s$=24, $T/a_t=256$}           & \multicolumn{1}{c}{$L/a_s$=24, $T/a_t=256$} \\
\hline
\hline
\multirow{4}{*}{[000] $A_1^+$ }
              & $\pi[000] K[000]$ $\pi[100] K[100]$   & $\pi[000] K[000]$ $\pi[100] K[100]$ & $\pi[000] K[000]$ $\pi[100] K[100]$ \\ 
              & $\pi[110] K[110]$ $\pi[111] K[111]$   &  & \\
              & $\eta[000] K[000]$ $\eta[100] K[100]$ & $\eta[000] K[000]$                   & $\eta[000] K[000]$\\
              & $\qquad 7\times q\bar{q}$             & $\qquad 18 \times q\bar{q}$     & $\qquad 5 \times q\bar{q}$     \\
\hline
\multirow{3}{*}{[000] $T_1^-$}
              & $\pi[100] K[100]$ $\pi[110] K[110]$   & $\pi[100] K[100]$             & $\pi[100] K[100]$ \\
              & $\eta[100] K[100]$                    &                               & \\	
              & $\qquad 11 \times q\bar{q}$           & $\qquad 44 \times q\bar{q}$   & $\qquad 8 \times q\bar{q}$     \\
\hline
\multirow{3}{*}{[000] $E^+$}
              & $\pi[100] K[100]$ $\pi[110] K[110]$    & $\pi[100] K[100]$             & $\pi[100] K[100]$ \\
              & $\eta[100] K[100]$                     &                               &\\	
              & $\qquad 22 \times q\bar{q}$            &  $\qquad 26 \times q\bar{q}$   & $\qquad 16 \times q\bar{q}$ \\
\hline
\multirow{6}{*}{[100] $A_1$}
              & $\pi[000] K[100]$ $\pi[100] K[000]$    & $\pi[000] K[100]$ $\pi[100] K[000]$    & $\pi[000] K[100]$ $\pi[100] K[000]$     \\ 
              & $\pi[100] K[110]$ $\pi[110] K[100]$    & $\pi[100] K[110]$ $\pi[110] K[100]$    & $\pi[100] K[110]$ $\pi[110] K[100]$\\
              & $\pi[110] K[111]$ $\pi[111] K[110]$    & \\
              & $\pi[100] K[200]$ $\pi[200] K[100]$    & \\
              & $\eta[100] K[000]$ $\eta[000] K[100]$  & $\eta[100] K[000]$ $\eta[000] K[100]$  & $\eta[100] K[000]$ $\eta[000] K[100]$\\
              & $\qquad 8\times q\bar{q}$     & $\qquad 14 \times q\bar{q}$                     & $\qquad 11\times q\bar{q}$     \\  
\hline
\multirow{6}{*}{[110] $A_1$}
              & $\pi[000] K[110]$ $\pi[100] K[100]$    & $\pi[000] K[110]$ $\pi[100] K[100]$    & $\pi[000] K[110]$ $\pi[100] K[100]$  \\ 
              & $\pi[110] K[000]$ $\pi[110] K[110]$    & $\pi[110] K[000]$                      & $\pi[110] K[000]$  \\
              & $\pi[100] K[111]$ $\pi[111] K[100]$    &                                        &\\   
              & $\eta[110] K[000]$ $\eta[100] K[100]$  & $\eta[110] K[000]$ $\eta[100] K[100]$  & $\eta[110] K[000]$ $\eta[100] K[100]$ \\
              & $\eta[000] K[110]$                     & $\eta[000] K[110]$                     & $\eta[000] K[110]$\\
              & $\qquad 9\times q\bar{q}$              & $\qquad 24 \times q\bar{q}$                   & $\qquad 12 \times q\bar{q}$     \\  
\hline
\multirow{5}{*}{[111] $A_1$}
              & $\pi[000] K[111]$ $\pi[100] K[110]$    & $\pi[000] K[111]$ $\pi[100] K[110]$    & $\pi[000] K[111]$ $\pi[100] K[110]$  \\ 
              & $\pi[110] K[100]$ $\pi[111] K[000]$    & $\pi[110] K[100]$ $\pi[111] K[000]$    & $\pi[110] K[100]$ $\pi[111] K[000]$  \\
              & $\eta[111] K[000]$ $\eta[110] K[100]$  & $\eta[111] K[000]$ $\eta[000] K[111]$  & $\eta[111] K[000]$ $\eta[000] K[111]$ \\
              & $\eta[100] K[110]$ $\eta[000] K[111]$  & \\
              & $\qquad 21\times q\bar{q}$             & $\qquad 12 \times q\bar{q}$            & $\qquad 13 \times q\bar{q}$  \\
\hline
\multirow{4}{*}{[200] $A_1$}
              & $\pi[100] K[100]$ $\pi[000] K[200]$     & $\pi[100] K[100]$ $\pi[000] K[200]$   & $\pi[100] K[100]$ $\pi[000] K[200]$ \\ 
              & $\pi[110] K[110]$ $\pi[200] K[000]$     & $\pi[110] K[110]$ $\pi[200] K[000]$   & $\pi[110] K[110]$ $\pi[200] K[000]$\\
              & $\eta[100] K[100]$                      & $\eta[100] K[100]$ $\eta[200] K[000]$ & $\eta[100] K[100]$ $\eta[200] K[000]$\\
              & $\qquad 23\times q\bar{q}$              & $\qquad 9 \times q\bar{q}$            &  $ \qquad 24 \times q\bar{q}$      \\   
         
\hline
\multirow{4}{*}{[100] $E_2$}
              & $\pi[100] K[110]$ $\pi[110] K[100]$     & $\pi[100] K[110]$ $\pi[110] K[100]$   & $\pi[100] K[110]$ $\pi[110] K[100]$     \\ 
              & $\pi[110] K[111]$ $\pi[111] K[110]$     &                                        & \\
              & $\eta[110] K[100]$ $\eta[100] K[110]$   &                                        &       \\
              & $\qquad 33\times q\bar{q}$              & $\qquad 16 \times q\bar{q}$     & $\qquad 39\times q\bar{q}$     \\  
\hline
\multirow{4}{*}{[110] $B_1$}  
              & $\pi[100] K[100]$ $\pi[110] K[110]$     & $\pi[100] K[100]$                     & $\pi[100] K[100]$   \\ 
              & $\pi[110] K[200]$ $\pi[200] K[110]$     &                                        &\\   
              & $\eta[100] K[100]$ $\eta[110] K[110]$   & $\eta[100] K[100]$                    & $\eta[100] K[100]$  \\
              & $\qquad 30\times q\bar{q}$              & $\qquad 16 \times q\bar{q}$           & $\qquad 14 \times q\bar{q}$     \\  
\hline
\multirow{3}{*}{[110] $B_2$}
              & $\pi[100] K[111]$ $\pi[111] K[100]$    &                                        &   \\ 
              & $\pi[110] K[110]$                      &                                        &    \\
              & $\qquad 28\times q\bar{q}$             & $\qquad 52 \times q\bar{q}$            & $\qquad 16 \times q\bar{q}$     \\  
\hline
\multirow{3}{*}{[111] $E_2$}
              & $\pi[100] K[110]$ $\pi[110] K[100]$    & $\pi[100] K[110]$ $\pi[110] K[100]$   &  $\pi[100] K[110]$ $\pi[110] K[100]$  \\
              & $\eta[100] K[110]$ $\eta[110] K[100]$  &                                        & \\
              & $\qquad 33\times q\bar{q}$             & $\qquad 20 \times q\bar{q}$            & $\qquad 16 \times q\bar{q}$  \\
\hline
\multirow{3}{*}{[200] $E_2$}  
              & $\pi[110] K[110]$ $\pi[111] K[111]$    & $\pi[110] K[110]$                     &  $\pi[110] K[110]$  \\ 
              & $\eta[110] K[110]$                     &                                       &                  \\ 
              & $\qquad 33\times q\bar{q}$             &  $\qquad 16 \times q\bar{q}$          &  $ \qquad 39 \times q\bar{q}$                  
\end{tabular}
\caption{The operators used in obtaining the energy levels shown in Fig.~\ref{fig_all_spec}. Numbers in square brackets denote the type of momentum, where $\vec{p} = [ijk] = \frac{2\pi}{L}(i,j,k)$. Momentum directions are summed in constructing meson-meson-like operators as described in Ref.~\cite{Dudek:2012gj}.  By $q\bar{q}$ we refer to constructions of the form $\bar{\psi}\Gamma D...D\psi$ that resemble a single-meson structure. The operators used at the largest pion mass are listed in Ref.~\cite{Wilson:2014cna}. $a_t m_\ell$ and $a_t m_s$ appear in the lattice action from Refs.~\cite{Edwards:2008ja,Lin:2008pr}.}
\label{tab:ops}
\end{table}

\end{document}